\documentclass{llncs}  

\usepackage{graphicx} 
\usepackage[T1]{fontenc}
\usepackage{amsmath}
\usepackage[english]{babel}
\usepackage{array,tabularx}
\usepackage{booktabs} 
\usepackage{makeidx}  
\usepackage{xparse}
\usepackage{moredefs}
\usepackage{lips}
\usepackage{epstopdf} 
\usepackage{color}
\usepackage{csquotes}
\usepackage{xcolor}
\usepackage{times}
\usepackage[hidelinks]{hyperref}
\usepackage[nolist,nohyperlinks]{acronym}
\usepackage{comment}
\usepackage{paralist}
\usepackage{cleveref}
\usepackage{wrapfig}
\usepackage{multirow}
\usepackage{microtype}
\usepackage{listings}
\usepackage[super]{nth}

\usepackage{float}
\floatstyle{plaintop}
\restylefloat{table}

\usepackage[style=base,labelfont=bf,labelsep=period,tableposition=top,font=small]{caption}
\makeatletter
\renewcommand\@biblabel[1]{#1.}
\makeatother
\addto\captionsenglish{}

\usepackage{fancyhdr}
\fancypagestyle{WI_footer}{\fancyhf{}\fancyfoot[L]{\color{black}\scriptsize 17th International Conference on Wirtschaftsinformatik\\February 2022, Nürnberg, Germany}}

\usepackage{cite}

\crefformat{footnote}{#2\footnotemark[#1]#3}
\expandafter\def\expandafter\UrlBreaks\expandafter{\UrlBreaks
  \do\a\do\b\do\c\do\d\do\e\do\f\do\g\do\h\do\i\do\j%
  \do\k\do\l\do\m\do\n\do\o\do\p\do\q\do\r\do\s\do\t%
  \do\u\do\v\do\w\do\x\do\y\do\z\do\A\do\B\do\C\do\D%
  \do\E\do\F\do\G\do\H\do\I\do\J\do\K\do\L\do\M\do\N%
  \do\O\do\P\do\Q\do\R\do\S\do\T\do\U\do\V\do\W\do\X%
  \do\Y\do\Z}

\newcolumntype{L}[1]{>{\raggedright\arraybackslash}p{#1}}   
\newcolumntype{C}[1]{>{\centering\arraybackslash}p{#1}}     
\newcolumntype{R}[1]{>{\raggedleft\arraybackslash}p{#1}}    

\begin{document}
\frontmatter          

\mainmatter              

\title{Supporting Domain Data Selection in \\ Data-Enhanced Process Models}


\author{Jonas Cremerius  \and
Mathias Weske}

\institute{Hasso-Plattner-Institute, University of Potsdam, Potsdam, Germany\\
\email{\{jonas.cremerius, mathias.weske\}@hpi.de}}

\maketitle
\setcounter{footnote}{0}

\begin{abstract} 
Process mining bridges the gap between process management and data science by discovering process models using event logs derived from real-world data. Besides mandatory event attributes, additional attributes can be part of an event representing domain data, such as human resources and costs. Data-enhanced process models provide a visualization of domain data associated to process activities directly in the process model, allowing to monitor the actual values of domain data in the form of event attribute aggregations. However, event logs can have so many attributes that it is difficult to decide, which one is of interest to observe throughout the process. This paper introduces three mechanisms to support domain data selection, allowing process analysts and domain experts to progressively reach their information of interest. We applied the proposed technique on the MIMIC-IV real-world data set on hospitalizations in the US.\\

\end{abstract}

\thispagestyle{WI_footer}


\section{Introduction}

Business organizations are confronted with the challenge of dealing with more and more data provided by their information systems. Those systems store real-world process execution data, from which process models can be discovered using process discovery techniques. The discovered process models represent the actual execution of processes, which allows confirming expected process behaviour and might reveal unknown insights about the processes~\cite{PMAction}. 

So far, process models can be analysed using events and their ordering. As information systems provide additional domain data stored as event attributes, data-enhanced process models allow a data-based exploration of processes. In data-enhanced process models, domain data can be displayed at activities, where an event attribute was used or modified, allowing to observe how domain data is influenced by the process~\cite{DEP}.

In comparison to other work in the literature, data-enhanced process models allow monitoring the development of the actual values of domain data during process execution in the process model, as, for example, the laboratory values during hospitalization might change throughout the process. This can help to check, for example, clinical guidelines, which make statements about the data behind events, such as that elevated concentrations of circulating cardiac troponins are detected in the vast majority of patients with acute heart failure~\cite{HF_guide}. 

However, an event might have many event attributes. For example, an event representing a stay in a hospital department can include over 100 laboratory values, medication data, imaging results, vital signs, and more \cite{lab_values, EHR}. Choosing the event attribute for observation in the data-enhanced process model is difficult, if not impossible, with so many event attributes available. Does the value of an event attribute change throughout the process, or is it only used at one process step? If it changes, which activities modify the event attribute, and what is the degree of variability? Therefore, the process analysts and domain experts are confronted with information overloading when faced with a long list of available event attributes. 

Information highlighting is a technique to control information overloading aiming to reduce confusion, and to increase accuracy and efficiency of understanding the information presented~\cite{Highlighting}. 
This paper provides an approach, in which highlighting is used to support the information selection process of domain data for illustration in data-enhanced process models.
Concretely, we introduce three mechanisms helping process analysts and domain experts to observe the most interesting event attributes from event logs depending on their type, variability, and process characteristic.


This contribution follows a design science approach consisting of five steps before publication~\cite{ds}. The first step, \textit{Problem Identification and Motivation}, was presented in the first section. After that, Section II describes the second step, \textit{Objectives of a Solution}, by illustrating related work facing the stated problem and current solutions in other fields dealing with the problem. Section III and IV present the \textit{Design and Development} step, introducing preliminaries and the approach. Section V covers the last two steps, \textit{Demonstration} and \textit{Evaluation}, by applying the approach to the MIMIC-IV real-world data set on hospitalizations. We discuss the approach and its limitations in Section VI before the paper is concluded in Section VII.

\section{Related Work}

Even though process discovery mainly focuses on control flow, the visualization of domain data has been researched as well.

In data-enhanced process models, domain data can be monitored in the form of event attribute aggregations, enabling a data-based exploration of the process. There, the process analysts and domain experts choose the activity and the event attribute for aggregation from a list, which is then visualized at the respective activity of the process model~\cite{DEP}. Another example is the multi-perspective process explorer, where the value distribution of selected domain data can be observed at each activity in the form of histograms~\cite{process-explorer}. In data-aware heuristic mining, domain data is chosen from a list to identify conditional dependency relations or decision rules~\cite{data-aware-heuristic}. 

These contributions have in common, that these leave the users with the choice of the event attribute of interest. This can be a problem when facing a lot of available event attributes, leading to information overloading.

The approach to support process analysts and domain experts on finding event attributes of interest is related to the topic of highlighting in information visualization, which deals with the problem of information overloading. In particular, the interaction control for information retrieval should provide mechanisms allowing users progressively to reach their interested information~\cite{Highlighting}.

One mechanism to highlight event attributes is related to the classification of event attributes. Event attributes have been classified in several ways for different analysis purposes. For instance, \cite{class_ML} classifies event attributes into controllable and non-controllable, where the first one can be altered during a process execution and the other cannot. Another contribution classifies domain data according to a medical standard (openEHR) to find dependencies within and between these categories \cite{class_EHR}. 

The other mechanism of highlighting refers to the discipline of change detection. Change detection is widely spread in image detection and involves measuring how the attributes of an area have changed between two dates \cite{change_detection_def}. It can be used to identify, for example, flooding, earthquakes, or urban sprawl. Change detection can also be applied for an arbitrary amount of images. In \cite{change_detection_CV}, the coefficient of variation (CV) is used to measure the change across N images of a particular area. Considering a construction site, the CV can measure the degree of variability during the process of construction.

The approach in this paper addresses the problem of finding event attributes of interest for analysis in a data-enhanced process model, which has not been tackled so far to our knowledge. The objective is to reduce the cognitive load of process analysts and domain experts, allowing them to observe the most interesting information in the process. To achieve that, this contribution brings the disciplines of information highlighting, event attribute classification, and change detection together to provide an artefact supporting domain data selection in data-enhanced process models. 

\section{Preliminaries}
\label{Sec_3}
This paper builds on data-enhanced process models, which are mined from an event log. An event log consists of sequences of events, which are grouped into traces. An event can have an arbitrary number of additional event attributes. The following definition is based on~\cite{correlation}.

\subsubsection{Definition 1 (Event log, Trace, and Event)}

Let $V$ be the universe of all possible values and $E_{A}$ be the universe of event attributes. An event $e$ is a mapping of event attributes to values, such as $e \in E_{A} \to V$. The universe of events is defined as $E_{U} = E_{A} \to V$. A trace $t \in {E_{U}}^\star$ is a sequence of events, and $T = {E_{U}}^\star$ represents the respective universe of traces. An event log $L$ is a multi-set of traces, so $L \in M(T)$, where, given any set X, $M(X)$ is the set of all possible multi-sets over $X$.
Normally, an event represents an activity which is conducted within a certain case at a given timestamp. These are treated as regular event attributes in this contribution, so we assume activity, case, and timestamp. The event instances of a given trace are ordered by their timestamp and have the same case. If an event $e \in E_{U}$ has no value assigned to an event attribute $e_{At} \in E_{A}$, it is denoted as $e(e_{At}) = \bot$. We further assume, that the data type of one event attribute is always the same for all events and that each event is unique. As every event is unique, we say, that two traces $t1, t2 \in L$ are identical, if the events in the traces have the same activity ordering.

In process mining, process models are discovered from event logs, which are defined in the following.

\subsubsection{Definition 2 (Process Model)} Given an event log $L \in M(T)$ and a process discovery algorithm $\alpha$ taking $L$ as its input, then $\alpha(L)$ generates a process model $P = (N, E)$. $N$ is the set of process activities and represents the events in the event log, such that given an event $e \in E_{U}$ and any event attribute $e_{At} \in E_{A}$ representing the process activity in the log $L$, $e(e_{At}) \in N$. In the following, we assume, that $e(activity)$ is used as the process activity. $E$ is a set of directed edges, such that $E \subseteq N \times N$. Gateways or other model-specific elements are not considered in this definition, as data-enhanced process models modify the activities of process models only.

To make event attributes presentable at activities in process models, event attribute aggregations take all event attributes related to an activity and provide different operations for aggregation. The definition is based on~\cite{DEP}.

\subsubsection{Definition 3 (Event Attribute Aggregation)} An event attribute aggregation is a tuple $e_{AA} = (a, e_{At}, E_{AV}, \lambda_{A})$ consisting of

\begin{itemize}
  \item an activity $a \in N$ for which the event attributes are extracted
  \item an event attribute $e_{At} \in E_A$, which is aggregated
  \item a multi-set $E_{AV}$ with all event attribute values to be aggregated 
  \item an aggregation function $\lambda_{A}$ taking all event attribute values in $E_{AV}$ and aggregating them to a rational number
\end{itemize}
Given an event log $L$, the desired event attribute values can be extracted by iterating through each trace $t \in L$, including the contained events. In the following, we use the normal set definition notation for multi-sets, as we specify beforehand, if the set is a multi-set:
\begin{equation}
    E_{AV} := \{e(e_{At}) \in V \mid e(activity) = a, e \in t, t \in L \}
\end{equation}

The event attribute aggregations can be linked to a process model, resulting in a data-enhanced process model. The definition is based on \cite{DEP} as well.

\subsubsection{Definition 4 (Data Enhanced Process Model)}
Let $E_{AA}$ be the universe of event attribute aggregations. An event attribute aggregation $e_{AA}$ is a tuple, as defined in Definition 3, such that $e_{AA} \in E_{AA}$. Given a process model $P = (N, E)$, a data-enhanced process model $DEP$ enhances each activity $N$ by a set of event attribute aggregations, resulting in $N_{DEP} \subseteq (N \times E_{AA})$. The data-enhanced model is then a tuple of data-enhanced activities $N_{DEP}$ and edges $E$, such that $DEP = (N_{DEP}, E)$.\\

Measuring and comparing the degree of variability between different characteristics, such as different event attributes, is achieved by using the coefficient of variation (CV), which is also known as the relative standard deviation.

\subsubsection{Definition 5 (Coefficient of Variation (CV))}

Given the mean ($\mu$) and standard deviation $(\sigma)$, the CV is defined as follows~\cite{CV_def}:

\begin{center}
   $CV(\%) = \frac{\sigma}{\mu} * 100$ 
\end{center}

Dividing the standard deviation by the mean creates a relative measure, allowing to compare the degree of variability between different characteristics. It is undefined for $\mu = 0$, which might be the case when data is normalized or negative values are in the data.

\section{Approach}
\label{Sec_4}

In this contribution, we help the user to find an event attribute $e_{At} \in E_{A}$ for a selected activity $act \in dom(N_{DEP})$ to be represented as an event attribute aggregation $e_{AA} \in E_{AA}$. Thus, we provide support for creating the tuple $(act, e_{AA}) \subseteq N_{DEP}$, which is the foundation of data-enhanced process models $DEP = (N_{DEP}, E)$. To achieve that, three mechanisms are provided to reduce the amount of selectable domain data. The reason why we provide mechanisms instead of recommending a certain event attribute is that the preferences on which attribute might be interesting can vary among different domain experts and process analysts. Some might be interested in looking at a specific attribute at a given process step, while others are interested in observing the development of an event attribute throughout the process. The mechanisms are related to the purpose of data-enhanced process models, which aim at understanding the process behaviour of the actual values of domain data.

Therefore, we start by separating categorical from quantitative event attributes. Then, we define process characteristics of domain data, which classifies domain data according to their process behaviour. Finally, we provide further filtering of domain data with a certain process characteristic based on the CV.

\subsubsection{Differentiation between categorical and quantitative event attributes}

The idea behind separating categorical from quantitative event attributes is to get a first overview about the domain data in the event log, which is a well established data pre-processing step~\cite{attribute_classification}. Additionally, data-enhanced process models provide different forms of aggregations for both types of variables~\cite{DEP}.

One approach to separate both classes is to calculate the number of unique values and divide this by the number of all values in a data-set, which was chosen based on current literature~\cite{attribute_classification}.  
For any given event attribute $e_{At} \in E_{A}$, we say that the set $e_{Unique}$ contains all unique values and the multi-set $e_{Values}$ includes all values of $e_{At}$ in the event log. Dividing $|e_{Unique}|$ by $|e_{Values}|$ results in a ratio $cf$, which can be used to classify the event attributes. This ratio can be compared to a threshold $th$, where a low value indicates a categorical and a high value a quantitative type. For example, if all values of an event attribute are unique, $cf$ is 1, indicating a quantitative attribute. If three of 100 values are unique, $cf$ is $0.03$, which is likely to be categorical.
Choosing the threshold is dependent on the view of what a categorical variable is. The threshold should have the property that variables with a lot of unique values have a $cf \gg th$ and variables with a comprehensible amount of categories have a $cf \ll th$.

Having the domain data classified into categorical and quantitative, the process characteristics of them are assessed in the following.

\subsubsection{Classifying event attributes according to their process characteristics}

In process mining, process analysts try to understand the process behaviour of events in the event log. The process behaviour is graphically represented by the control flow and optionally gateways, which bring structure into the events. It can be seen, if events with the same activity name occur multiple times or just once during a process instance. Thus, events in an event log have different process characteristics. 

In this contribution, we propose to give event attributes a process characteristic as well. An event attribute might be used only once in a process. Another one might be used multiple times and evolves throughout the process.

Therefore, the following classification according to process characteristics of event attributes is suggested. Two features are provided to classify the process characteristic of domain data. The first feature is the total number of activities in which the event attribute is set. Formally, it is defined as follows: 

Given an event attribute $e_{At} \in E_{A}$ in an event log $L$, the set $e_{Act}$ represents all activities in which the event attribute is used.

\begin{equation}
       e_{Act} :=  \{e(activity) \in V \mid e(e_{At}) \neq \bot, e \in t, t \in L\}
\end{equation}

The cardinality $|e_{Act}|$ represents the number of activities in which the respective event attribute occurs.

The second feature, $e_{AvgTrace}$, is the average number of times, the event attribute occurs per trace. This feature is important to answer the question, whether an event attribute is used once or multiple times during a process instance. Hence, it can be evaluated, whether an event attribute evolves throughout the process. This feature is calculated by iterating through each trace in which the event attribute is used at least once. In these traces, the number of events having the respective event attribute are counted. 

Formally, this is defined as follows, given an event attribute $e_{At} \in E_{A}$ and an event log $L \in M(T)$: First, the event log is filtered, so that only traces are included which use the event attribute at least once. 

\begin{equation} \label{eq:3}
       L' = \{t \in L \mid (\exists e \in t) [e(e_{At}) \neq \bot] \}
\end{equation}

Then, the average number of events using the event attribute can be calculated:

\begin{equation}
       e_{AvgTrace} = \frac{\sum_{t \in L'} \sum_{e \in t} [e(e_{At}) \neq \bot]}{|L'|}
\end{equation}

The equation above uses the Iverson bracket, which evaluates to 1, if the condition is true and otherwise to 0. Hence, if the event attribute is not $\bot$, 1 is returned and added to the sum \cite{iverson}.

Three different process characteristics are defined in this contribution based on the previously defined features $|e_{Act}|$ and $e_{AvgTrace}$. The first process characteristic describes event attributes which do not change throughout the process and occur at only one activity in all process instances. That characteristic is called \emph{static}. For instance, the admission location of a patient in a hospital could be recorded only at the hospital admission. 

The second one includes event attributes which do not change throughout the process, but occur at different activities in the process. Looking at a logistic process, where one of multiple means of transportations can be chosen, the attribute transfer duration could occur at the activities "Transport via ship" or "Transport via truck", but always only once per process instance. That one is called \emph{semi-dynamic}. 

The last process characteristic describes domain data which changes throughout the process and occurs at multiple activities. That one is called \emph{dynamic}. An example could be the state of an order, which is created at the beginning by one activity and finished/cancelled by other activities at the end of the process. In the following, the presented process characteristics are formalized:

\[
pc (e_{At}) = \left.
\begin{cases}
    static, &  |e_{Act}| = 1, e_{AvgTrace} = 1 \\
    semi-dynamic, &  |e_{Act}| > 1, e_{AvgTrace} = 1 \\
    dynamic, &  |e_{Act}| \geq 1, e_{AvgTrace} > 1
\end{cases}
\right\}
\]

It should be noted, that $e_{AvgTrace}$ is a rational and $|e_{Act}|$ is a natural number. $e_{AvgTrace}$ cannot be $< 1$, because we look only at traces including the event attribute at least once.
Considering loops, the condition for the dynamic process characteristic, that $|e_{Act}| \geq 1$ is important. If there is a loop of a single process activity changing a unique event attribute, $|e_{Act}|$ is 1, but the event attribute occurs multiple times in a trace. 

This mechanism gives process analysts and domain experts an overview of the process behaviour of domain data, where they know in advance, if an event attribute at a given activity can be used for monitoring throughout the process or is exclusively available for the activity. 

Looking at \emph{dynamic} event attributes, the degree of variability can be specified based on the CV introduced previously.

\subsubsection{Filtering dynamic event attributes according to their degree of variability} 

\emph{Dynamic} event attributes have the particular property of evolving throughout the process. However, some event attributes might change through the process execution and some might not. Measuring the degree of variability of domain data would enable process analysts and domain experts to filter event attributes according to their degree of variability, identifying event attributes which are sensitive to the process. Thus, they can decide if they desire to analyse event attributes, which remain constant throughout the process or have a high degree of variability.  

In this paper, the CV is proposed to measure the degree of variability of domain data. As mentioned in related work, the CV has been used in the field of change detection in images, taking a sequence of pixels and measuring their change over time \cite{change_detection_CV}. Inspired by this methodology, this contribution uses the CV to measure the degree of variability of domain data.

\begin{table}[htb]
\caption{Example event log}
\centering
\begin{tabular}{|c|c|c|c|c|}
\hline
Case ID & Activity & Timestamp & \shortstack{Glucose \\Value} & \shortstack{Creatinine \\Value} \\
\hline
1 & Admit to hospital & 1 & 140 & 0.7 \\
\hline
1 & Treat in Medical Ward & 2 & 200 & 0.7 \\
\hline
1 & Discharge Patient & 3 & 120 & 0.8 \\
\hline
2 & Admit to hospital & 1 & 135 & 0.6 \\
\hline
2 & Treat in ICU & 2 & 175 & 0.6 \\
\hline
2 & Discharge Patient & 3 & 110 & 0.7 \\
\hline
\end{tabular}
\label{tab:example_log}
\end{table}

The reason for choosing the CV is its ability to identify variability and its independency to the scale of a variable, which allows comparing variables of different scales \cite{CV_def}. To demonstrate that, an example event log is presented in Table \ref{tab:example_log}, showing two traces with two additional event attributes representing laboratory values in a hospital.
The example event log contains two event attributes which are occurring in multiple activities and multiple times in a trace. Therefore, these are classified as \emph{dynamic}. Hence, these can be further filtered by their degree of variability. The laboratory values of glucose and creatinine have different scales, which makes a comparison by their variance (41 mg/dL vs. 0.05 mg/dL for case 1) useless. However, the CV as a relative measure results in a comparable number for both variables (27\% vs. 7\% for case 1). Thus, the glucose value has a higher degree of variability than the creatinine value in case 1, which can also be confirmed by looking at the values in the event log. The same can be calculated for case 2, where the CV for the glucose value is 22\%. Taking the average of both traces results in an average CV of 24.5\% for the glucose value. 

The method to calculate the CV of a given event attribute $e_{At} \in E_{A}$ for an event log $L \in M(T)$ is formalized in two steps.

The first step is to calculate the CV of $e_{At}$ for each trace $t \in L'$ representing a process instance, where $L'$ is the event log including traces where the event attribute occurs at least once, as defined in equation \ref{eq:3}. For that, the function $CV(t) = t \to CV$ takes the event attribute values of a given trace and returns the CV of this sequence. The multi-set $CV_{eAt}$ contains the CV of all traces in the event log $L'$.

\begin{equation}
   CV_{eAt} = \{ CV(t): t \in L' \}
\end{equation}

Having the CV of each trace for an event attribute $e_{At}$, this contribution defines the degree of variation of an event attribute $DegVar_{eAt}$ as the average CV of all traces in the event log $L'$.

\begin{equation}
  DegVar_{eAt} = \frac{\sum_{CV \in CV_{eAt}} CV}{|CV_{eAt}|}
\end{equation}

The CV is only useful for event attributes having a ratio scale~\cite{CV_def}. Ratio scale variables are non-negative, which is not the case for all quantitative variables. For example, the temperature is a quantitative variable, but its values can be negative. This kind of variable is on a so-called interval ratio.

One reason why the CV is unsuitable for negative values is the danger of having a mean close to zero, which would distort the CV, as the standard deviation is divided by the mean to calculate it~\cite{CV_book}.
Dealing with negative event attributes can be solved by shifting all data points by adding all numbers with the absolute of the minimum value, such that the minimum becomes zero and the other numbers become positive. As this contribution uses the CV for detecting variability, the actual values do not matter, but only the distance between them. Therefore, this type of normalization is suitable for dealing with negative data. For example, the sequence $[-2, 2, 4]$ is transformed to $[0, 4, 6]$ by adding the absolute minimum value (2) to all numbers. 

Additionally, if interval scale variables can be converted to different scales, such as \textdegree C to \textdegree F, the CV can be manipulated. That is due to the fact, that the ratios between two temperature measurements in \textdegree C or \textdegree F are not defined, as 40 \textdegree C is not twice as warm as 20 \textdegree C. Therefore, the ratios change when converting the values, and with that the degree of variability measured by the CV. This can be solved by converting the variables to a ratio scale, as, for example, the temperature can be measured in \textdegree K, which is on a ratio scale. 

Looking at categorical data, it is not possible to calculate the mean and standard deviation of all categorical event attributes, as some of them might be encoded as strings. However, a technique is proposed to transform categories into numerical values to calculate the CV. The transformation is based on~\cite{CV_book}: The so-called mapping method maps the categories of an event attribute $e_{At}$ with discrete values from 1 to n, with n being the total number of categories. First, the frequency of all possible values of $e_{At}$ is obtained, where the values are sorted according to their frequency. Then, the mapping is performed by assigning the discrete values from 1 to n to the categories of $e_{At}$, corresponding to their position in the sorted frequency list. Using the mapped values, a feasible numerical representation for CV calculation is available.


Next, this approach is evaluated on a real-world healthcare data set, provided by the MIMIC-IV database.

\section{Evaluation}
\label{Sec_5}

The proposed approach was implemented in ProM as an extension of the data-enhanced process model implementation, which is based on the inductive visual miner\footnote{\url{https://github.com/jcremerius/Data-Enhanced-Process-Models}}. The implementation is available on GitHub, which can be downloaded\footnote{\url{https://github.com/jcremerius/Supporting-Domain-Data-Selection-in-DEP}}. The relevance of this approach is illustrated in a medical environment, where an event log was extracted from the Medical Information Mart for Intensive Care IV (MIMIC-IV) database. The reason for creating an event log from this database is to illustrate an example, where event attributes are available which are measured throughout the whole process, such as laboratory values or vital signs. 

\subsection{Dataset}

MIMIC-IV is a relational database representing the whole patient journey through a hospital, including procedures performed, medications given, laboratory values taken, image analysis conducted, and more. Its purpose is to support research in healthcare and is therefore publicly available \cite{MIMIC}. 

The event log extracted from MIMIC-IV incorporates a high-level process, where the patient journey through different hospital departments, such as emergency department, medical ward or intensive care unit (ICU), is stored. The event log contains 1050 hospital process instances with 8646 events of heart failure patients treated in the ICU.

An event can have up to 67 event attributes, including laboratory values, vital signs, status scores, and demographic information. The number of available event attributes could be much higher in such a hospital process (imaging data, medications, procedures, and more), where this event log focusses on the most prevalent values from the MIMIC-IV database only. Thus, it contains 52 laboratory value measurements, 6 demographic data points, and 9 ICU related measurements (vital signs and status scores). 

\subsection{Results}
\begin{figure}[h]
    \centering
    \includegraphics[width=4.6cm]{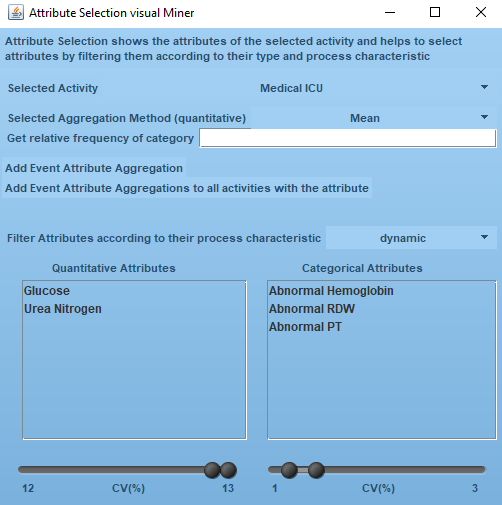}
    \caption{Attribute selection view showing dynamic event attributes}
    \label{fig:selection_1}
\end{figure}

We applied the approach explained in Section \ref{Sec_4} on the dataset introduced above. Thus, the event attributes were separated according to their quantitative or categorical type. Then, these were classified according to their process characteristic and the CV was calculated for the \emph{dynamic} event attributes. This information can then be used to find the event attribute of interest to be displayed in a data-enhanced process model, which is illustrated in Fig. \ref{fig:selection_1}.

At the top, the view shows the selected activity and the means of aggregation, whereas at the bottom two lists are illustrated, which show the separation between quantitative and categorical event attributes. The selection list above the two lists allows filtering according to the three process characteristics. The two sliders at the bottom allow further filtering of \emph{dynamic} domain data according to the CV(\%). 

The event attributes are separated by their quantitative or categorical type. The quantitative variables include the actual values of the measurements conducted, whereas the categorical event attributes include booleans indicating an abnormal laboratory value or strings describing the insurance or admission location of a patient. We used a threshold $th$ of $0.05$ to separate domain data, where all quantitative and categorical data types were identified correctly. 

With the help of the proposed filtering techniques, the number of event attributes to choose from can be reduced significantly. The two boxes at the bottom of Fig. \ref{fig:selection_1} show the result of an arbitrary filtering process, where event attributes having a \emph{dynamic} process characteristic are shown, which are in the specified CV range illustrated by the sliders. 

The activity "Medical ICU" has 26 quantitative and 26 categorical laboratory value measurements which are \emph{dynamic}. It has none \emph{static} and 11 \emph{semi-dynamic} event attributes. Laboratory values are measured in completely different scales. Nevertheless, these can be compared by their degree of variability, where the laboratory value of ``Glucose'' has a relatively high degree of variability with a CV of 12.7\%. On the other hand, the categorical variable ``Abnormal Hemoglobin'' has a relatively low degree of variability in the process, with a CV of 2.7\%. Thus, the view can be modified, so that there is only a comprehensible amount of event attributes left for selection. When selecting the desired event attribute, the preferred aggregation can be chosen, which can then be either displayed solely at the selected activity or at all activities using the event attribute.

\begin{figure}
  \noindent\makebox[\textwidth]{%
  \includegraphics[width=1.4\textwidth]{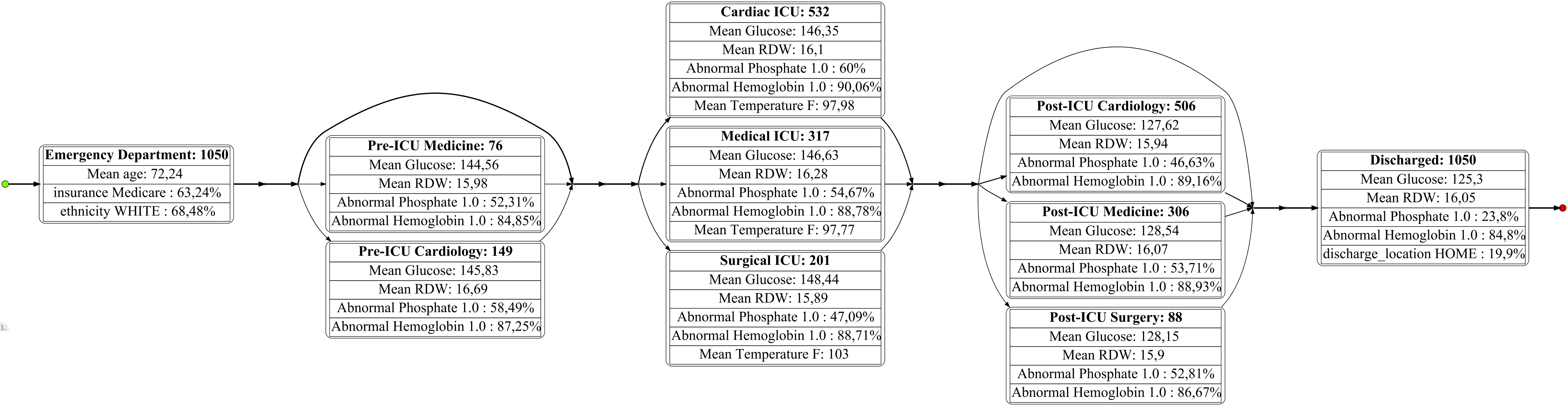}}
  \caption{Data-enhanced process model with event attributes selected based on their process behaviour. RDW: Red Cell Distribution Width}
  \label{fig:dep_2}
\end{figure}

Fig. \ref{fig:dep_2} illustrates a data-enhanced process model with event attributes selected based on the information filtering. We decided to show the mean of the quantitative event attributes to represent the whole patient group. It can be seen, that the laboratory value of ``Glucose'' varies throughout the process, as it is between 144 mg/dL and 148 mg/dL during treatment and then drops to 125 mg/dL after treatment in the ``Post-ICU'' and ``Discharged'' activities. On the other hand, the laboratory value of ``Red Cell Distribution Width (RDW)'' barely changes during process execution and remains at 16\% for the patients with heart failure, which is also represented by a CV of 1.3\%.

The same can be observed for the categorical event attributes, where ``Abnormal Phosphate'' (CV: 14.4\%) varies and ``Abnormal Hemoglobin'' (CV: 2.7\%) remains similar throughout the process. Additionally, the activity ``Emergency Department'' has three \emph{static} event attributes including age, insurance, and ethnicity, which are only available for this activity. That is the case for the ``discharge\_location'' at the activity ``Discharged'' as well. The ``Temperature F'' is a \emph{semi-dynamic} event attribute, which is exclusively measured at the three ICU wards, where heart failure patients in this process stay only once during their hospitalization. 
This discussion shows, that domain data selection by expert users is significantly improved by the techniques proposed in this paper, because it is already known during the filtering process, whether domain data is used at the selected activity only or evolves throughout the process, and, if so, to which degree event attributes are sensitive to the process.

\section{Discussion}
\label{Sec_6}
This paper suggests a method to classify domain data according to their process characteristics and measuring their degree of variability, which is used to filter event attributes. With that, data-based exploration in data-enhanced process models should be improved by reducing information overloading when process analysts together with domain experts look for domain data of interest.

Supporting information filtering based on process characteristics and the degree of variability goes hand-in-hand with the process analysis possibilities of data-enhanced process models. These include \emph{Monitoring event attribute values} and \emph{Comparison of process activities}, suggesting observing, whether an event attribute is changing throughout the process or identifying differences between event attribute values in different activities \cite{DEP}. With the approach presented in this contribution, it can be determined, if an event attribute is changing and, if so, at what degree in comparison to others. By displaying that event attribute as an event attribute aggregation in a data-enhanced process model, the activities causing the variability can be identified.

This contribution suggests using the CV as a measure of variability, which was chosen based on current research dealing with change detection. The evaluation shows promising results to use the CV as a measure to compare the degree of variability between different event attributes. However, the need of converting domain data being on an interval ratio or categorical still leaves room for improvement. 

It should be noted, that this kind of analysis is of explorative nature and not creating any causal evidence. The whole concept of data-enhanced process models aims at enhancing the ability of process exploration in process mining by supporting the visualization of event attributes, which gives a first impression on the behaviour of domain data in the process.

\section{Conclusion and Future Work}

This contribution researches methods to filter domain data, supporting domain experts and process analysts to find the information of interest. The proposed mechanisms allow reducing the number of event attributes to choose from to a comprehensive amount by filtering them according to their process characteristics and their degree of variability. 

Future work could focus on enhancing the methodology and evaluation. For example, further filtering mechanisms can be developed, such as a comparison of semi-dynamic attributes regarding their difference between activities. The differentiation between categorical and quantitative event attributes could also be improved by applying further methods, such as machine learning techniques, to automatically identify the most suitable threshold for an event log. Additional evaluation by talking to domain experts or applying the approach to more use cases could be conducted as well. 

\clearpage

\bibliographystyle{splncs04}
\bibliography{./literature.bib}

\end{document}